\documentclass[12pt,letterpaper]{revtex4}
\usepackage{amsmath}
\usepackage{blindtext}
\usepackage{extarrows}
\usepackage{braket}
\usepackage{pdfpages}
\usepackage{graphicx}
\usepackage{caption}
\usepackage{float}
\graphicspath{ {images/} }
\usepackage{amsmath,graphicx,color,xcolor,epsfig}
\usepackage{epstopdf}
\usepackage{hyperref}
\begin{document}

\begin{titlepage}

\begin{center}
		
{\bf\Large\boldmath Global quantum discord and von Neumann entropy in multipartite two-level atomic systems}\\[15mm]
\setlength {\baselineskip}{0.2in}
{\large  M. Ibrahim, M. Usman, Khalid Khan}\\[5mm]

{\it Department of Physics, Quaid-i-Azam University, Islamabad 45320, Pakistan.}\\[5mm]
		
\end{center}
	
{\bf Abstract}\\[5mm] 
\setlength{\baselineskip}{0.2in} 
We have computed the global quantum discord and von Neumann entropy of multipartite two-level atomic systems interacting with a single-mode Fock field. We use Tavis-Cumming model. We have explored how quantum correlations and quantum entanglement evolve with time in such systems. The quantum system is prepared initially in a mixed state and different parameters are varied to see how they affect the information processing in the system. The dynamical character of the global quantum discord and von Neumann entropy show an interplay between classical and non-classical correlations. Photons in this model play an important role to assist the global quantum discord and von Neumann entropy and we observed that the effects of the field on the global quantum discord and von Neumann entropy reside in the time evolution of the system indicating that both atom and field states have become entangled. The global quantum discord is assisted in a non-linear fashion with the number of photons in the system. The global quantum discord and von Neumann entropy show linear behavior with each other in the dynamics of the system. The effects of intrinsic decoherence on the dynamics of the global quantum discord and von Neumann entropy are also studied. We have extrapolated the results for a large photon number on the system. We have studied the effect of the change in the size of the system on the maximum value of global quantum discord and von Neumann entropy and we have estimated the scaling coefficients for this behavior.
\end{titlepage}
\section{INTRODUCTION}
Quantum computing requires the involvement of quantum systems consisting of two-level subsystems \cite{bennett2000quantum}. The success of quantum computation depends on the controlled evolution and precise measurements of these quantum systems. Various phenomena in quantum information and computation such as entanglement and quantum correlation using two-level atoms and spin models are explored and understood \cite{amico2008entanglement,osborne2002entanglement,vidal2003entanglement,osterloh2002scaling,yang2019enhancing}. However, quantifying entanglement and quantum correlations for the multipartite systems remains a theoretical challenge. Quantum correlations are more meaningful to study in many-body quantum systems as compared to bipartite entanglement \cite{amico2008entanglement,legeza2004quantum}. However these features show similar behavior for pure states \cite{rulli2011global,campbell2013global}.\\
Entanglement is an important resource in the field of quantum information processing and quantum computing \cite{nielsen2002quantum,RevModPhys.81.865}. Bipartite entanglement is well studied in various aspects \cite{wootters,walborn2006,peres1996,vidal2002,poon2007,ryu2008,yu2010,miranowicz2008}. For instance, the entanglement of mixed states in two qubits system is characterized by Wootters \cite{wootters} and can be experimentally measured \cite{walborn2006}. As for the multipartite systems, tripartite entanglement has triggered considerable interests in the study to increase the security in quantum cryptography \cite{hayashi2006} and in the efficiency of quantum cloning \cite{zhang2000,groblacher2006}. Tripartite entanglement \cite{leon2009,casagrande2009,huang2010,militello2011}, processing of quantum information \cite{braunstein1988,horodecki1994} and entropy\cite{bennett1996,brassard1996} have been studied extensively. For a bipartite pure state, the von Neumann entropy (VNE) of the reduced density matrix of either subsystem is a good and unique measure\cite{brassard1996,popescu1997}. Relative entropy of the system has been proposed as a measure of entanglement and effectively used for the mixed states \cite{vedral997,vedral1998}.\\
It has been identified that entanglement is not the only parameter to judge quantumness of a system \cite{henderson2001classical} by introducing a suitable tool to measure quantum correlations, known as quantum discord, presented by Ollivier and Zurek \cite{ollivier2001quantum}. Quantum discord and its dynamics has been extensively explored in spin models and cavity quantum electrodynamics \cite{dillenschneider2008quantum,werlang2010quantum,maziero2012long,rossatto2011nonclassical,berrada2012classical,he2011sudden,chen2010quantum,campbell2011propagation}. Quantum correlations have also been thought as a useful measurement in quantum evolution under decoherence \cite{maziero2009classical,maziero2010system,ferraro2010almost,mazzola2010sudden,huang2012different} and a resource for quantum computation \cite{datta2008quantum} have the property that almost all types of quantum states have non-vanishing quantum discord \cite{ferraro2010almost}.\\
A multipartite version of quantum discord, called global quantum discord (GQD) has been studied and derived by Rulli and Sarandy \cite{rulli2011global}. The non-classical states in multipartite systems are studied by Saguia et. al. \cite{saguia2011witnessing} which provides a sufficient condition for non-classicality of the system. The GQD in the thermal Ising model has been investigated in the reference \cite{campbell2011global}.\\
We present a study of the GQD and VNE in the $N$ two-level atoms coupled with the single-mode field in a Fock state. We investigate the effect of different parameters that are present in the model, such as intrinsic decoherence, parameters in the initial atomic state, and the number of photons present in the system, on the GQD and VNE. The paper is arranged as follows; In section II we briefly describe the model and the initial state of the system and we present multipartite quantum correlations. In section III, we study the behavior of the GQD and VNE. The behavior of both the quantifiers is analyzed regarding the field and with various possible initial states. The effects of intrinsic decoherence in the system are also studied. The results are studied with the increase in the size of two-level atomic systems. In the last section, we conclude our findings.\\
\section{Global Quantum Discord for the multipartite quantum two-level system}
\subsection{The Model}
We explore multipartite two-level atomic systems interacing with Fock field. The Hamiltonian of the system we study is,\\
\begin{equation}
\hat{H}=\dfrac{\omega_{0}}{2} \sum_{i=1}^{N} \hat{\sigma}_{i}^{z}+\omega \hat{a}^{\dagger}\hat{a}  +g \sum_{i=1}^{N}(\hat{a}\hat{\sigma}_{i}^{+}+\hat{a}^{\dagger}\hat{\sigma}_{i}^{-})
\label{equation 1}
\end{equation}
where $\hat{\sigma}_{i}^{z}$ is the $z$-component of Pauli matrix whereas $\hat{a}_i$ ($\hat{a}_{i}^{\dagger}$) are creation (annihilation) operators and $\hat{\sigma}_{i}^{+}$ ($\hat{\sigma}_{i}^{-}$) are atomic raising (lowering) operators. The frequencies associated with two-level atomic transition and the field are denoted by $\omega_0$ and $\omega$ respectively, whereas $g$ is the coupling parameter  between atom and the field. The first term in Hamiltonian corresponds to the atoms, second term is associated with field and the third term in Hamiltonian describes the atom field interaction under rotating wave approximation.  The time evolution of the system under the Markovian approximation is given by \cite{milburn1991intrinsic},
\begin{equation}
\dot{\hat{\rho}}(t)=-i[\hat{H},\hat{\rho}(t)]-\frac{\gamma}{2} [\hat{H},[\hat{H},\hat{\rho}(t)]]
\label{equation 2}
\end{equation}
where $\rho$ is the density matrix and $\gamma$ is the coefficient of intrinsic decoherence. For $\gamma\rightarrow 0$, Eq. (\ref{equation 2}) reduces to the standard von Neumann equation describing the Schrodinger equation. The formal solution of Eq. (\ref{equation 2}) is given by,
\begin{equation}
\hat{\rho}(t)=\sum_{k=0}^{\infty}\frac{(\gamma t)^{k}}{k!}\hat{M}^{k}(t)\hat{\rho}(0)\hat{M}^{k\dagger}(t),
\label{equation 3}
\end{equation}
with
\begin{equation}
\hat{M}^k(t)=\hat{H}^k\exp(-i\hat{H}t)\exp(-\gamma t \hat{H}^2 /2),
\label{e4}
\end{equation}
where $\hat{\rho}(0)$ is the initial state of the system.\\

We assume atoms and field ae initially uncoupled, thus we prepare initial state of the system $\hat{\rho}_{AF}(0)$ as a product state
\begin{equation}
\hat{\rho}_{AF}(0)=[(1-p)\ket{\psi} \bra{\psi}+p\ket{g_1 g_2 ... g_N} \bra{g_1 g_2 ... g_N}]\otimes \ket{n} \bra{n},
\label{initial state}
\end{equation}
where $\ket{\psi}=\cos(\alpha) \ket{g_1 g_2 ... g_N} + \sin(\alpha) \ket{e_1 e_2 ... e_N}$,  $\ket{g_i}$ and $\ket{e_i}$ are the ground and  excited states of the two-level atoms respectively. p corresponds to statistical probability, $0\leq p \leq 1$, $\alpha$ is associated with super position of two level system, $0\leq\alpha\leq\pi$ and $\ket{n}$ is the field state. In our model, the combined system i.e. the atoms and field form the set of allowable basis states $\{\ket{\psi_i}\}$ of the system given as,
\begin{equation}
\begin{split}
\{\ket{\psi_i}\}=&\ket{g_1,g_2,g_3,...g_N,n+N},
\ket{e_1,g_2,g_3,...g_{N},n+N-1},
\ket{e_1,e_2,g_3,...g_{N},n+N-2},\dots\\
&\ket{e_1,e_2,e_3,...e_{N},n}
\end{split}
\label{eq.5}
\end{equation}
The $ij^{th}$ matrix element of the Hamiltonian in the allowed basis is $\bra{\xi_{i}}\hat{H}\ket{\xi_{j}}$ with $\ket{\xi_{i}}=\bigotimes_{l=1}^{N}\ket{s}_{l}$ where $s$ represents the ground and excited state of the $l^{th}$ two-level atomic system, the basis are $\ket{0}=\begin{bmatrix} 1 \\
0 \end{bmatrix}$ and  $\ket{1}=\begin{bmatrix} 0 \\
1 \end{bmatrix}$. Now for the two two-level atomic system, the matrix elements of the Hamiltonian in the allowable basis are,
\begin{equation}
 \left(
\begin{array}{cccc}
0 & g \sqrt{n+1} & g \sqrt{n+1} & 0 \\
g \sqrt{n+1} & 0 & 0 & g \sqrt{n+2} \\
g \sqrt{n+1} & 0 & 0 & g \sqrt{n+2} \\
0 & g \sqrt{n+2} & g \sqrt{n+2} & 0 \\
\end{array}
\label{eq.6}
\right).
\end{equation}
The matrix elements for the large-N systems, more specifically for three, four and five two-level atomic system, are given in the appendix. Eq. (\ref{equation 3}) and Eq. (\ref{initial state}) in the allowable basis $\{\ket{\psi_{i}}\}$ are used to obtain the final state of the system at time $t$,
\begin{equation}
\hat{\rho}_{AF}(t)=\sum_{i,j;i\neq j}^{N}\exp[-\dfrac{\gamma t}{2}(E_i-E_j)^2-i(E_i-E_j)t]\times \bra{\psi_{i}}\hat{\rho}(0)\ket{\psi_{j}}\ket{\psi_i}\bra{\psi_j}
\label{fullstate}
\end{equation}
where $E_{i}$, $E_{j}$ are the eigenvalues of the Hamiltonian in the states $\{\ket{\psi_{i}}\}$. The final state of the atomic system is obtained after taking the trace over the field i.e. $\hat{\rho}(t)=Tr_{F}\big[\hat{\rho}_{AF}(t)\big]$.
\subsection{Multipartite Quantum Correlations} 
 For a bipartite system, composed of two subsystems A and B, quantum discord $D^{A\rightarrow B}$ is the difference between quantum mutual information $I(\rho)$ and classical correlations $J(\rho)$, minimized over the whole set of orthogonal projective measurements $\hat{\varPi}$ performed on the subsystem $B$ \cite{campbell2013global}, mathematically
\begin{equation}
D^{A\rightarrow B}(\rho_{ab})= \underset{\{\hat{\varPi}_{B}^{j}\}}{\text{min}} [I(\rho_{(AB)})-J(\rho_{(AB)})_{\{\hat{\varPi}_{B}^{j}\}} ]
\label{equation 4}
\end{equation}
Using the projective measurements on the system, the GQD for $N$-party system can be written as
\begin{equation}
GQD(\rho)=\underset{\{\hat{\varPi}^{j}\}}{\text{min}} [S(\rho_{T}||{\hat{\varPi}}(\rho_{T}))-\sum_{j=1}^{N} S(\rho_{j}||{\hat{\varPi}}_{j}(\rho_{j}))]
\label{equation 5}
\end{equation}
where $\rho_T$ is the total state and $\rho_j$ is the reduced state of subsystem $j$. The terms $S(\rho_{1}||\rho_{2})=Tr[\rho_{1}\ln \rho_{1}]-Tr[\rho_{1}\ln \rho_{2}]$ are the relative entropies between two generic states $\rho_{1}$ and $\rho_{2}$. The idea of bipartite quantum discord has been generalized to multipartite qauantum correlation, also called the GQD in the reference \cite{Campbell2013}. In order to calculate the global quantum correlations we can use an alternate form of the GQD \cite{Campbell2013} given below,
\begin{equation}
GQD(\rho_{T})=\underset{\{\varPi^{k}\}}{\text{min}}\biggl\{\sum_{j=1}^{N}\sum_{l=0}^{1}\tilde{\rho}_{j}^{ll}\log_{2}\tilde{\rho}_{j}^{ll}-\sum_{k=0}^{2^{N}-1}\tilde{\rho}_{T}^{kk}\log_{2}\tilde{\rho}_{T}^{kk}\biggr\}+\sum_{j=1}^{N}S(\rho_j)-S(\rho_T)
\label{equation 10}
\end{equation}
where $\tilde{\rho}_{T}^{kk}=\bra{k}\hat{R}^{\dagger}\rho_{T}\hat{R}\ket{k}$ and $\tilde{\rho}_{j}^{ll}=\bra{l}\hat{R}^{\dagger}\rho_{j}\hat{R}\ket{l}$, and $\hat{\varPi}^k= \hat{R}\ket{k}\bra{k}\hat{R}^{\dagger}$ are the multi-qubit projective operators. $S(\rho_{j})=-Tr[\rho_{j}\log_{2}\rho_{j}]$ and $S(\rho_{T})=-Tr[\rho_{T}\log_{2}\rho_{T}]$ are the von Neumann entropies of the subsystem $j$ and the total system respectively. Here $\{\ket{k}\}$ are the eigenstates of $\bigotimes _{j=1} ^{N} \hat{\sigma} _{j} ^{z}$ and $\hat{R}$ is a local multi-qubit rotation operator acting on the $j$th qubit, expressed as $\hat{R}=\bigotimes _{j=1} ^{N} \hat{R} _ {j} (\theta_{j},\phi_j)$ with $\hat{R} _ {j} (\theta_{j},\phi_j)=\cos \theta _{j} \hat{1}+i \sin \theta _{j} \cos \phi_{j} \hat{\sigma}_{y} +i \sin \theta _{j} \sin \phi_{j} \hat{\sigma}_{x}$. Expression of the GQD given in Eq. (\ref{equation 10}) significantly reduces the computational efforts required to evaluate the quantum correlations by using the defined projective operators. Alongside GQD, the dynamics of VNE ($S(\rho)=-Tr[\rho\log_{2}\rho]$) is also computed and compared with GQD.\\
\section{Results and discussion}
In this section, we explore the dynamics of the GQD and VNE of the system with an initially mixed state given by Eq. (\ref{initial state}). The dynamical behavior of both quantifiers, the GQD and VNE are numerically evaluated for the different number of photons $n$, intrinsic decoherence $\gamma$, and the mixing parameter $\alpha$.\\
\subsubsection{\textit{Global quantum discord and von Neumann entropy in two and five two-level atomic system interacting with single mode Fock field}}
\begin{figure}[H]
	\centering
	\includegraphics[width=7in]{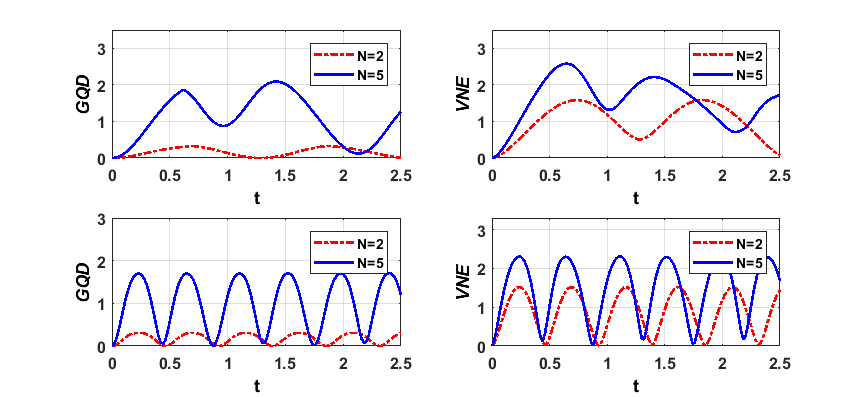}
	\caption{(color online) The GQD and VNE are plotted as a function of scaled time for $N=2$ and $N=5$ two-level atomic systems. The plots in the upper panel are for photon number $n=0$ and in the lower panel are for $n=10$.  All data is for $\gamma=0$, $p=0$ and $\alpha=0$.}
	\label{fig:1}
\end{figure}

Fig. (\ref{fig:1}) shows the results for the time evolution of the GQD and VNE for the system with two ($N=2$) and five ($N=5$) two-level atoms. For $N=2$ a periodicity in gradual increase and decrease of the GQD and von Neumann entropy is seen in the system. It shows a rise and fall of quantum correlations. There is no abrupt vanishing of quantum correlations in the system. Such behavior has also been seen by Fanchini et. al. and Werlang et. al. \cite{fanchini2010non,werlang2009robustness} for a two-qubit system where the sudden death of discord is not observed instead it remains periodic. The maximum value of VNE for $N=2$ suggests a higher degree of entanglement in the system whereas zero value indicates that all the atoms are in only one atomic state. As the size of the system increases with the addition of two-level atoms, the magnitudes of both GQD and VNE show an increasing behavior. Extrema of both the GQD and VNE does not appear at the same time especially for large-N systems where the atoms have more accessible atomic states. The dynamical behavior of these quantifiers suggests that the information can be retained in such systems that can be useful in quantum processing and quantum computation. Moreover, the effect of photon number in the system plays a significant role in the dynamics of the quantifiers. The behavior of the GQD and VNE for non-zero photon number $n$ is also shown in Fig. (\ref{fig:1}). Rather periodic behavior of both quantifiers is present and that has not been seen in zero photon case. This periodicity can be seen as the availability of the photons to cause atomic transitions in the system. We observe the effects of the field on the quantifiers, which is present even after tracing out the field from the final density matrix. Quantum systems composed of two-level atoms, and the environment with which they interact retains some information of the system \cite{pknight}. Therefore the effect of field resides in the time evolution of density matrix that ensures both atom and field states have become a mixed state. This periodic behavior in both quantifiers is seen for the larger-N system as well.\\
\subsubsection{\textit{Effect of mixing parameter $\alpha$}}
\begin{figure}[H]
	\centering
	\includegraphics[width=7in]{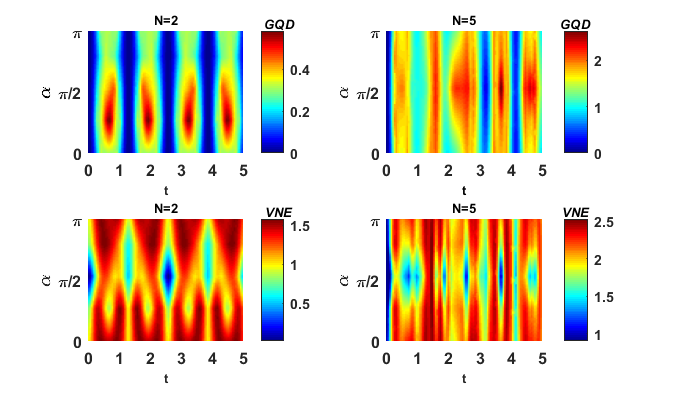}
	\caption{(color online) The density plots of GQD and VNE as a function of $\alpha$ and scaled time $t$ for $N=2$ and $N=5$ two-level atomic systems. All data is set for $\gamma=0$, $n=0$ and $p=0.5$.}
	\label{fig:3}
\end{figure}

In Fig. (\ref{fig:3}) density plots show the effect of mixing parameter $\alpha$ on the dynamics of the GQD and VNE. It is observed that certain values of mixing parameters assist the dynamics of GQD. For $N=2$, the system shows the maximum value of GQD around $\alpha=\pi/4$ that corresponds to the initially mixed state. For the large-N systems, the maximum value of GQD is observed around $\alpha=\pi/2$ that represents an initially mixed state prepared in an equal statistical mixture of the ground states and excited states. This behavior points out that the larger-N system favors the initial state prepared in $\alpha=\pi/2$. For the large-N systems, almost all states show non-zero quantum correlations.\\	
\subsubsection{\textit{Effect of photon number $n$ on the GQD}}
\begin{figure}[H]
	\centering
	\includegraphics[width=5.5in]{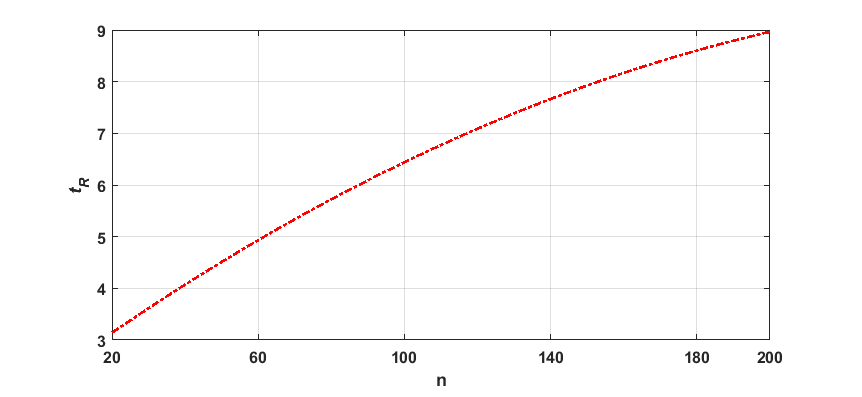}
	\caption{(color online) The number of maxima appearing in the GQD per unit time, $t_{R}$, is plotted against the number of photons $n$. $t_{R}$ increases in a rather non-linear fashion and increases if photons $n$ are increased. All data is for $\gamma=0$, $p=0$ and $\alpha=0$.}
	\label{fig:photonassis}
\end{figure}

\begin{figure}[H]
	\centering
	\includegraphics[width=7in]{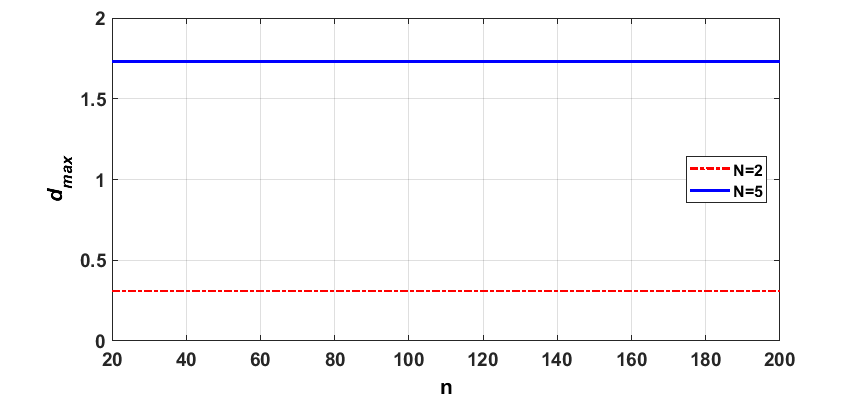}
	\caption{(color online) The maximum of the GQD of the system composed of two and five two-level atomic system is analyzed with the change in the photons $n$. There is no increase in the maximum value of quantum correlations in the system with $n$. All data is for $\gamma=0$, $p=0$ and $\alpha=0$.}
	\label{fig:photoneffectondiscord}
\end{figure}

\begin{figure}[H]
	\centering
	\includegraphics[width=7in]{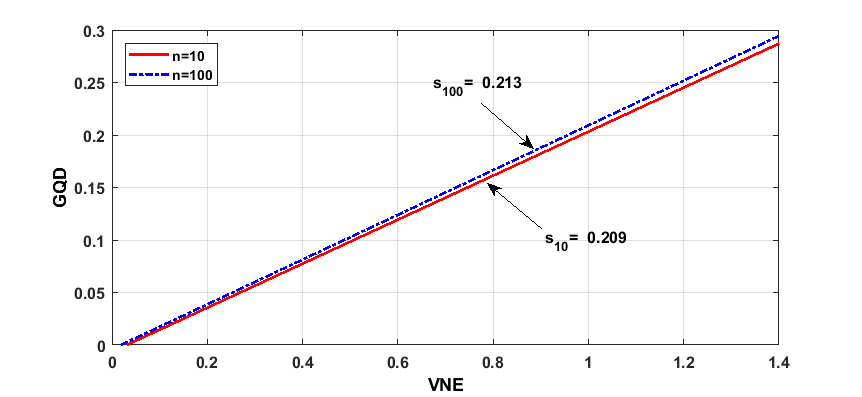}
	\caption{(color online) The slopes of the lines remain almost the same as the photons in the system are increased. Both quantifies change with almost same magnitude as the photons are changed in the system. All data is for $N=2$, $\gamma=0$, $p=0$ and $\alpha=0$.}
	\label{fig:vneslope}
\end{figure}

In Fig. (\ref{fig:photonassis}), number of maxima appearing per unit time against photon number $n$ is plotted. Photons in the system play an important role to assist and boost the correlations. The effect of increasing the photons in the system has two important consequences: firstly, increasing the photons support the quantum correlation in the system by reducing the revival time of the correlations, and secondly, the photons do not raise the maximum value of quantum correlations in that system. Furthermore, the increase in the number of revivals in the unit time $t$, defined by $m_{R}$, as shown in Fig. (\ref{fig:photonassis}) tends to increase non-linearly as the photons $n$ in the system are increased.\\
The maximum value of the GQD which is defined by $d_{max}$, for a system is plotted against the photons number $n$ in Fig. (\ref{fig:photoneffectondiscord}). The maximum value of the GQD in the system does not change as the number of photons are increased in the system. In Fig. (\ref{fig:vneslope}), the GQD is plotted against VNE for $n=10$ and $n=100$. As shown in Fig. (\ref{fig:vneslope}), the ratio between the GQD and VNE remains constant for two two-level atomic system. It is also observed that the slope between the GQD and VNE remains nearly the same as the photons are increased inside the system, as shown in Fig. (\ref{fig:vneslope}). There is a slight increase of slope as the photons are increased in the system. The effect of photon number $n$ is nearly the same with a slight change in the slope. Furthermore, according to Fig. (\ref{fig:1}), the behavior of $t_{R}$ to the change of the photons in the system is not affected in the larger-N systems and both the quantifies exhibits similar behavior.\\
\subsubsection{\textit{The dynamics of GQD and VNE in the presence of intrinsic decoherence}}
\begin{figure}[H]
	\centering
	\includegraphics[width=7in]{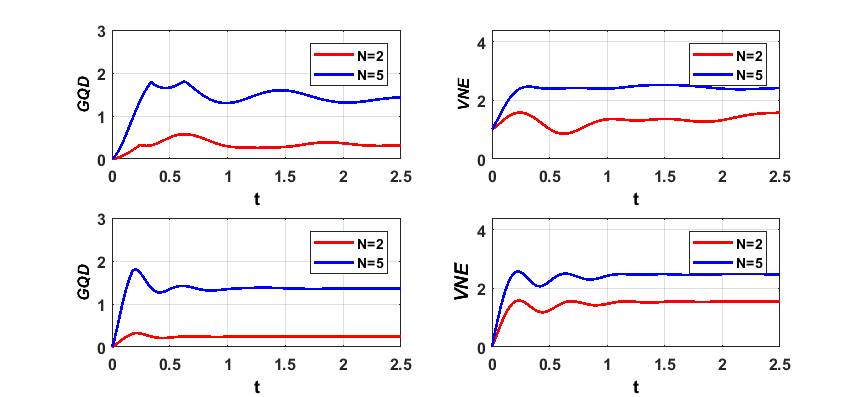}
	\caption{(color online) The GQD and VNE are plotted as function of scaled time for $N=2$ and $N=5$ two-level atomic systems for non-zero intrinsic decoherence. The plots in the upper panel are for photon number $n=0$ and in the lower panel are for $n=10$. All data is for $\gamma=0.05$, $\alpha=\pi/4$ and $p=0$.}
	\label{fig:4}
\end{figure}

\begin{figure}[H]
	\centering
	\includegraphics[width=6in]{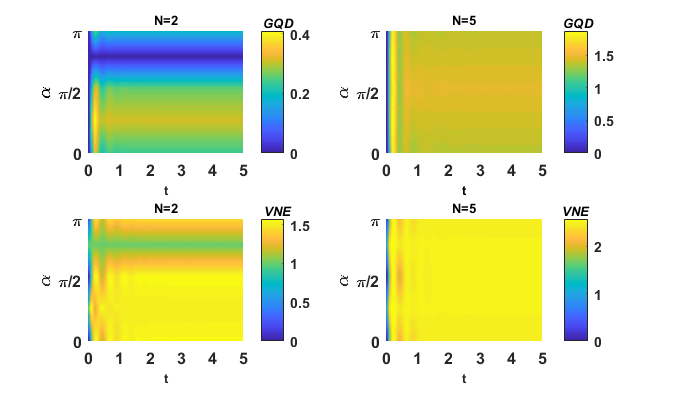}
	\caption{(color online) The density plots of GQD and VNE as a function of $\alpha$ and scaled time $t$ for $N=2$ and $N=5$ two-level atomic systems. All data is set for $\gamma=0.05$, $n=10$ and $p=0$.}
	\label{fig:6}
\end{figure}

The effect of intrinsic decoherence on the dynamics of the quantifiers for two and five two-level atoms is shown in Fig. (\ref{fig:4}). For the zero photon case, the dynamics of the GQD and VNE increase to a maximum value around $t\sim 0.5$. The maximum value of the GQD and VNE increases with the increase in the number of two-levels atoms $N$. The effects of the number of photons $n=10$ on the dynamics of the GQD and VNE are shown in lower panel of Fig. (\ref{fig:4}) lower panel. In the multiphoton case, the system gets smoother dynamical behavior in both quantifiers, as compared to zero photon case. The density plots in Fig. (\ref{fig:6}) show the GQD and VNE dynamics in the system with respect to $\alpha$. For $N=2$, quantum correlations are robust and vanishing for $\alpha=\pi/4$ and $\alpha=3\pi/4$, respectively.  It is observed that for two two-level atomic systems, the values of $\alpha$ have prominent effects on the dynamics of the GQD and VNE, whereas, for higher values of $N$, the system dynamics show less dependence on the mixing parameter $\alpha$.\\
\subsubsection{\textit{Purity versus the GQD and VNE}}
\begin{figure}[H]
	\centering
	\includegraphics[width=7in]{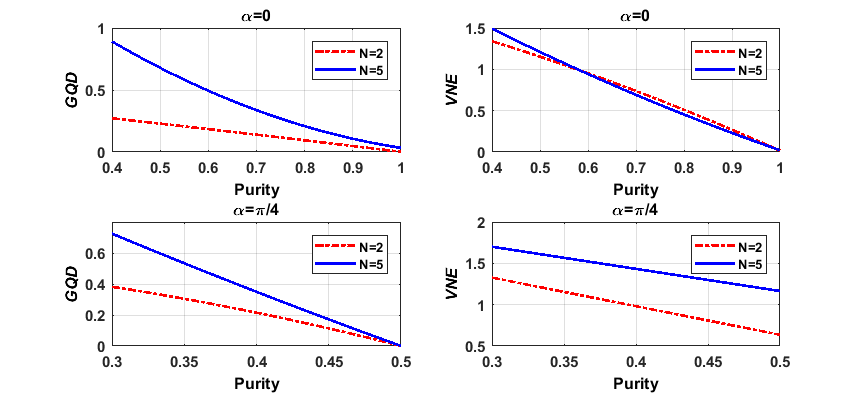}
	\caption{(color online) The variation in the correlations of the system is studied with respect to the purity of the system with two different initial states. All data is set for $\gamma=0$, and $p=0$.}
	\label{purity}
\end{figure}

The statistical mixture of the system as represented by density matrix $\hat{\rho}(t)$ can be quantified by measuring the quantum purity, which is defined as Tr$\rho(t)^{2}$. For a pure quantum state, purity is 1 i.e. Tr$\rho(t)^{2}=1$ whereas for a mixed state Tr$\rho(t)^{2}<1$. For an entangled state, the purity level of the subsystems of the multipartite system is always less than the purity of the full system represented in Eq. (\ref{fullstate}). The dynamical behavior between quantum purity and the GQD and VNE for two and five two-level atoms are plotted in Fig. (\ref{purity}). The plots are for two types of initial states, one is for $\alpha=0$ and the other for $\alpha=\pi/4$. For $\alpha=0$ when the system is allowed to evolve, both the GQD and VNE are non-zero for $\alpha=\pi/4$ state and zero for $\alpha=0$ state. In this case the quantum purity of the states satisfies $\frac{1}{d}\leq\textit{Tr}\rho^{2}\leq1$, where $d$ is the dimension of Hilbert space of multi two-level systems given as $2^{d}$. For an initial state at $\alpha=\pi/4$, the GQD is zero while VNE is not. We observe that the GQD and the purity of the state have an inverse relation. Similarly, the VNE and the purity of the state have an inverse relation with each other.\\

%\subsection{Effect of intrinsic decoherence on the dynamics of the system}
%
%
\subsubsection{\textit{Size of the two-level atomic system and the GQD and VNE}}
\begin{figure}[H]
	\centering
	\includegraphics[width=7in]{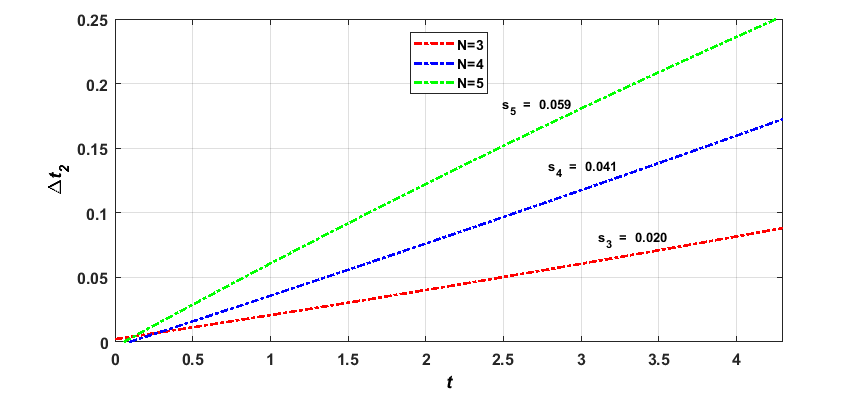}
	\caption{(color online) Shift in the compression time $\Delta t_{2}$ of the GQD is plotted for the three, four and five lwo-level atomic system with reference to the two two-level atomic system. There is an increase in the value of respective slope as system gets more two-level atoms. All data is for $n=10$, $\gamma=0$ and $\alpha=0$.}
	\label{compression}
\end{figure}

\begin{figure}[H]
	\centering
	\includegraphics[width=7in]{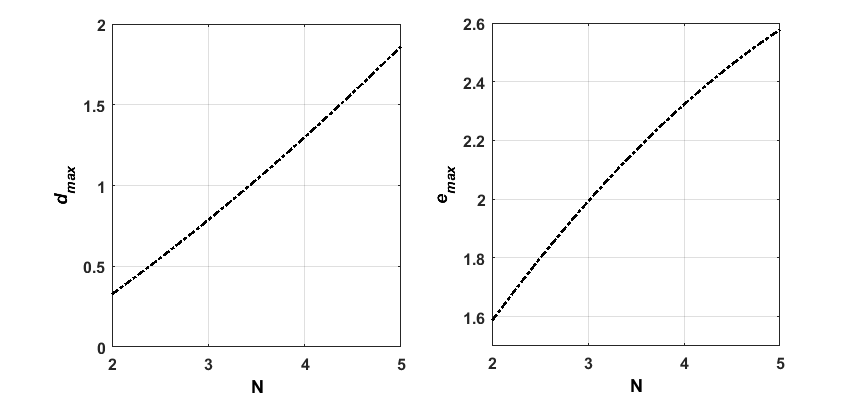}
	\caption{(color online) The change in maximum value of the GQD, denoted by $d_{max}$, and VNE, denoted by $e_{max}$, with the increase of each two-level atom in the system is plotted. The $d_{max}$ increases with the positive linear and quadratic curve fitting coefficients while $e_{max}$ increases with negative quadratic and positive linear coefficient. All data is for $\gamma=0$, $n=10$ and $p=0$.}
	\label{optimum}
\end{figure}

The general behavior of the GQD and VNE for a larger-N system is studied. From Fig. (\ref{fig:1}), it is seen that for the large N value, the magnitude of the GQD and VNE are increased and the first maximum of each system is shifted towards the zero of the time scale. Fig. (\ref{compression}) shows the shift factor of revivals of the GQD, denoted by $\Delta t_{2}$ of three, four, and five two-level atomic systems with the reference to the two two-level system. It is seen that with the addition of each two-level atom in the system, the slope of the line is increased by $0.02$ as compared to the two two-level atomic systems. The revival time of the GQD for the larger-N system will become more and more shifted towards the origin of the time scale and the system has quantum correlations reach to maximum value earlier in time. The maximum value of the GQD and VNE of the system achieved in the time evolution, denoted by $d_{max}$ and $e_{max}$ respectively, are plotted in Fig. (\ref{optimum}). Both $d_{max}$ and $e_{max}$ tend to increase with different non-linear fashion upon the increase in two-level atoms $N$. From Fig. (\ref{optimum}), the GQD varies with the quadratic ($a_{1}=0.025$) and linear ($b_{1}=0.331$) coefficient both positive. On the other hand the VNE, the coefficient with quadratic ($a_{2}=-0.037$) and linear ($b_{2}=0.955$) term has opposite signs.\\
\section{Conclusions}
In this paper we explored how global quantum discord and von neumann entropy evolve with time for multi two-level atomic systems interacting with the single mode Fock field. We computed the GQD and VNE for two, three, four and five two-level atomic system interacting with a single mode Fock field with and without intrinsic decoherence. We found that with increasing the size of the system (number of atoms), the GQD and VNE are enhanced which can be regarded as the change in the content of information. The photons assisted the quantum correlation by reducing the revival time of the quantifiers. The revivals in unit time had a non-linear behavior with the number of photons in the system. It was also observed that the maximum value of quantum correlations in the system did not change with the number of photons. The behavior of the quantifiers was also studied with different values of mixing parameters $\alpha$ with and without the intrinsic decoherence. The effect of the purity was analyzed and it showed an increase in the correlations that corresponded to less purity and a higher degree of mixing in the system. The effect of large-N on the quantifiers indicates that the GQD and VNE have different scaling behaviors.
\section{Appendix}
In this appendix we present the matrix elements for the large-N two-level atomic systems. For three two-level atomic system, the matrix elements are
\begin{equation*}
	\begin{split}
		&a_{1,2}=a_{1,3}=a_{1,5}=a_{2,1}=a_{3,1}=a_{5,1}=\sqrt{1+m},\\
		&a_{4,2}=a_{2,4}=a_{3,4}=a_{4,3}=a_{2,6}=a_{6,2}=a_{3,7}=a_{7,3}=a_{5,6}=a_{6,5}=a_{5,7}=a_{7,5}=\sqrt{2+m}\\
		&a_{4,8}=a_{8,4}=a_{6,8}=a_{8,6}=a_{7,8}=a_{8,7}=\sqrt{3+m}
	\end{split}
\end{equation*}
and the other elements are zero.\\
For four two-level atomic system, the matrix elements are
\begin{equation*}
	\begin{split}
		&a_{1,2}=a_{1,3}=a_{1,5}=a_{2,1}=a_{3,1}=a_{5,1}=a_{1,9}=a_{9,1}==\sqrt{1+m},\\
		&a_{4,2}=a_{2,4}=a_{3,4}=a_{4,3}=a_{2,6}=a_{6,2}=a_{3,7}=a_{7,3}=a_{5,6}=a_{6,5}=a_{5,7}=a_{7,5}=\\
		&a_{2,10}=a_{10,2}=a_{3,11}=a_{11,3}=a_{5,13}=a_{13,5}=a_{9,10}=a_{10,9}=a_{9,11}=a_{11,9}=\sqrt{2+m}\\
		&a_{4,8}=a_{8,4}=a_{6,8}=a_{8,6}=a_{7,8}=a_{8,7}=a_{4,12}=a_{12,4}=a_{6,14}=a_{14,6}=a_{7,15}=a_{15,7}=\\
		&a_{12,10}=a_{10,12}=a_{12,11}=a_{11,12}=a_{14,10}=a_{10,14}=a_{15,11}=a_{11,15}=a_{14,13}=\\
		&a_{13,14}=a_{15,13}=a_{13,15}=\sqrt{3+m}\\
		&a_{8,16}=a_{16,8}=a_{16,12}=a_{12,16}=a_{12,14}=a_{14,12}=a_{12,15}=a_{15,12}=\sqrt{4+m}
	\end{split}
\end{equation*}
and the other elements are zero.\\
For five two-level atomic system, the matrix elements are
\begin{equation*}
	\begin{split}
		&a_{1,2}=a_{1,3}=a_{1,5}=a_{2,1}=a_{3,1}=a_{5,1}=a_{1,9}=a_{9,1}=a_{17,1}=a_{1,17}=\sqrt{1+m},\\
		&a_{4,2}=a_{2,4}=a_{3,4}=a_{4,3}=a_{2,6}=a_{6,2}=a_{3,7}=a_{7,3}=a_{5,6}=a_{6,5}=a_{5,7}=a_{7,5}=\\
		&a_{2,10}=a_{10,2}=a_{3,11}=a_{11,3}=a_{5,13}=a_{13,5}=a_{9,10}=a_{10,9}=a_{9,11}=a_{11,9}=a_{18,2}=\\
		&a_{2,18}=a_{19,3}=a_{3,19}=a_{21,5}=a_{5,21}=a_{25,9}=a_{9,25}=a_{25,17}=a_{17,25}=a_{17,18}=\\
		&a_{18,17}=a_{17,19}=a_{19,17}=a_{21,17}=a_{17,21}=\sqrt{2+m}\\
		&a_{4,8}=a_{8,4}=a_{6,8}=a_{8,6}=a_{7,8}=a_{8,7}=a_{4,12}=a_{12,4}=a_{6,14}=a_{14,6}=a_{7,15}=a_{15,7}=\\
		&a_{12,10}=a_{10,12}=a_{12,11}=a_{11,12}=a_{14,10}=a_{10,14}=a_{15,11}=a_{11,15}=a_{14,13}=\\
		&a_{13,14}=a_{15,13}=a_{13,15}=a_{4,20}=a_{20,4}=a_{6,22}=a_{22,6}=a_{7,23}=a_{23,7}=\\
		&a_{10,26}=a_{26,10}=a_{11,27}=a_{27,11}=a_{13,29}=a_{29,13}=a_{26,18}=a_{18,26}=a_{27,19}=\\
		&a_{19,27}=a_{29,12}=a_{12,29}=a_{20,18}=a_{18,20}=a_{20,19}=a_{19,20}=a_{22,18}=a_{18,22}=a_{23,19}=\\
		&a_{19,23}=a_{22,21}=a_{21,22}=a_{23,21}=a_{21,23}=a_{26,25}=a_{25,26}=a_{27,25}=a_{25,27}=\\
		&a_{29,25}=a_{25,29}=\sqrt{3+m}\\
		&a_{8,16}=a_{16,8}=a_{16,12}=a_{12,16}=a_{12,14}=a_{14,12}=a_{12,15}=a_{15,12}=a_{24,8}=a_{8,24}=a_{28,12}=\\
		&a_{12,28}=a_{30,14}=a_{14,30}=a_{31,15}=a_{15,31}=a_{28,20}=a_{20,28}=a_{22,30}=a_{30,22}=a_{31,23}=\\
		&a_{23,31}=a_{24,20}=a_{20,24}=a_{24,22}=a_{22,24}=a_{24,23}=a_{23,24}=a_{28,26}=a_{26,28}=a_{28,27}=\\
		&a_{27,28}=a_{30,26}=a_{26,30}=a_{31,27}=a_{27,31}=a_{30,29}=a_{29,30}=a_{31,29}=a_{29,31}=\sqrt{4+m}\\
		&a_{16,32}=a_{32,16}=a_{24,32}=a_{32,24}=a_{32,28}=a_{28,32}=a_{32,30}=a_{30,32}=a_{32,31}=a_{31,32}=\sqrt{5+m}
	\end{split}
\end{equation*}
and the other elements are zero.\\

%\bibliographystyle{aipauth4-1}
%\bibliographystyle{plainnt}
%\bibliographystyle{plain}
%\bibliography{bibfile}

\end{document}